\documentclass[a4paper]{article}\usepackage[]{graphicx}\usepackage[]{color}
\makeatletter
\def\maxwidth{ %
  \ifdim\Gin@nat@width>\linewidth
    \linewidth
  \else
    \Gin@nat@width
  \fi
}
\makeatother

\definecolor{fgcolor}{rgb}{0.345, 0.345, 0.345}

\usepackage{framed}
\makeatletter
 {\par\unskip\endMakeFramed%
 \at@end@of@kframe}
\makeatother

\definecolor{shadecolor}{rgb}{.97, .97, .97}
\definecolor{messagecolor}{rgb}{0, 0, 0}
\definecolor{warningcolor}{rgb}{1, 0, 1}
\definecolor{errorcolor}{rgb}{1, 0, 0}
\newenvironment{knitrout}{}{} 

\usepackage{alltt}

\pdfoutput=1

\usepackage{alltt}
\usepackage{natbib}

\usepackage{float}

\usepackage{placeins}

\let\Oldsection\section
\renewcommand{\section}{\FloatBarrier\Oldsection}

\let\Oldsubsection\subsection
\renewcommand{\subsection}{\FloatBarrier\Oldsubsection}

\let\Oldsubsubsection\subsubsection
\renewcommand{\subsubsection}{\FloatBarrier\Oldsubsubsection}

\usepackage{mathptmx}      

\usepackage[]{graphicx}
\usepackage[]{color}
\usepackage[margin=25mm]{geometry}
\usepackage[hidelinks]{hyperref}
\usepackage{caption}

\usepackage{amsmath}
\usepackage{amssymb}

\IfFileExists{upquote.sty}{\usepackage{upquote}}{}
\IfFileExists{upquote.sty}{\usepackage{upquote}}{}
\begin{document}

\raggedbottom

\title{Plinko: Eliciting beliefs to build better models of statistical learning and mental model updating}

\author{Peter A.V. DiBerardino\thanks{Corresponding author}\\
  Department of Psychology\\
  University of Waterloo\\
  Waterloo, ON N2L 3G1\\
  \texttt{pavdiber@uwaterloo.ca}\and
  Alexandre L.S. Filipowicz\\
  Toyota Research Institute\\
  Los Altos, CA 94022\\
  \texttt{alsfilip@gmail.com}\and
  James Danckert\thanks{Contributed equally}\\
  Department of Psychology\\
  University of Waterloo\\
  Waterloo, ON N2L 3G1\\
  \texttt{jdancker@uwaterloo.ca}\and
  Britt Anderson\footnotemark[2]\\
  Department of Psychology\\
  University of Waterloo\\
  Waterloo, ON N2L 3G1\\
  \texttt{britt@uwaterloo.ca}
}

\maketitle

\begin{abstract}

  Prior beliefs are central to Bayesian accounts of cognition, but many of these accounts do not directly measure priors. More specifically, initial states of belief heavily influence how new information is assumed to be utilized when updating a particular model. Despite this, prior and posterior beliefs are either inferred from sequential participant actions or elicited through impoverished means. We had participants play a version of the game ``Plinko'', to first elicit individual participant priors in a theoretically agnostic manner. Subsequent learning and updating of participant beliefs was then directly measured. We show that participants hold a variety of priors that cluster around prototypical probability distributions that in turn influence learning. In follow-up experiments we show that participant priors are stable over time and that the ability to update beliefs is influenced by a simple environmental manipulation (i.e. a short break). This data reveals the importance of directly measuring participant beliefs rather than assuming or inferring them as has been widely done in the literature to date. The Plinko game provides a flexible and fecund means for examining statistical learning and mental model updating.

\bigskip
\begin{footnotesize}
\end{footnotesize}

\end{abstract}

\section{Introduction}
\label{intro}

Humans have a remarkable ability to learn complex statistical representations of the world \citep{saffran1996, NISSEN1987, Orban2745, turk2005}. We use this statistical information to build beliefs about our environment, sometimes called mental models, and update these beliefs when contingencies change \citep{Tenenbaum1279, fiser2010, JL2013}. The way we use such statistical information has become a key feature of many theories of learning and general cognition \citep{frost2019}, ranging from decision making \citep{fisk2002, Tenenbaum1279, summerfield2015} and language development \citep{saffran1996}, to measuring the cognitive consequences of brain damage \citep{danckert2012, filipowicz2016, stottinger2014, palminteri2012}.

At a minimum, mental models should contain representations of expected outcomes. These expectations ought to be based on prior experiences/beliefs. We argue that an appropriate method for examining mental models and statistical learning should be theoretically agnostic as to how the initial conditions of beliefs are represented. This is not trivial: research shows that the beliefs we use to interpret events can have a significant impact on decision making \citep{green2010, hock2005, lee2013, patrick2014, bianchi2020, stottinger2014priors}.

One approach is to examine response trends over multiple trials to characterize a participant’s expectations \citep{jueptner1997, jueptner1997II, NISSEN1987, robertson2001, toni1998, vulkan2000}. While such measures reveal how closely participants manage to match task contingencies, they give only limited information as to which beliefs were driving responses. As highlighted by \citet{stottinger2014}, data trends from individual responses alone can result from a number of different beliefs that may be unknown to the experimenter. 

Recent computational approaches have attempted to infer participant beliefs by modeling their behaviour \citep{nassar2010approximately, nassar2012rational, oreilly2013, mcguire2014, sepahvand2014, collins2012}. For example, Bayesian models of human learning represent participant beliefs as probability distributions, representing how likely events are to occur \citep{glaze2018, nassar2010approximately, nassar2012rational, oreilly2013, mcguire2014, Tenenbaum1279, griffiths2006optimal}. These distributions are then updated with each new observation, providing a dynamic representation of the way participant beliefs evolve throughout a task.

However, Bayesian models make important assumptions that should be accompanied by empirical evidence. Most prominently, the success of a Bayesian model depends largely on the prior, the distribution that represents the beliefs participants bring to a task before observing any information. These priors are often selected by the researchers themselves, and are assumed to be the same across participant groups. Critics have noted that these freedoms and assumptions in model design make Bayesian methods too flexible, rendering them essentially unfalsifiable \citep{bowers2012, jones2011}.

What is required, therefore, is a task that allows for flexible representations of participant beliefs without assuming a prior or its form. That is, explicitly collecting a prior from the individual rather than assuming a `one size fits all' approach. In the current article we present such a task. Based on the game ‘Plinko’, participants are given an intuitive environment in which, rather than make single responses, they draw distributions to indicate how likely they believe certain events are to occur. Our task affords the opportunity to collect idiosyncratic priors in a theoretically agnostic manner, before any evidence is presented to the participant, and track how these beliefs change as new evidence is presented.

Our task provides a more realistic representation of participants - as individuals who carry their own idiosyncratic beliefs (priors) or decision-making tendencies into a statistical learning task \citep{frost2019, franken2005}. In the case of `ecologically valid' tasks, these differences may be attributed to differences in accumulated knowledge \citep{siegelman2018linguistic}. In the case of more novel or abstract tasks, where previous life experience is less likely to directly inform optimal behaviour, individual differences in priors may still exist in the form of individual differences in information processing or perception. For example, someone who is optimistic may predict a higher proportion of `favourable' outcomes in a given task than someone who is pessimistic, even though they both have never before attempted the given task. Regardless, we neglect a crucial component of human statistical learning when we neglect individual differences in the initial conditions of belief.

Eliciting priors and trial-by-trial beliefs affords us the ability to invert the standard operating procedure of theoretical developments in belief updating. Treating participant beliefs as latent states to be inferred by participant actions requires candidate models of belief to be compared. If none of the candidate models correctly capture participants' latent beliefs, they are either doomed to fail or be wrongly adopted. Plinko instead provides an explicit theoretically agnostic measure of beliefs whose data can be used to form the appropriate theory.

There are existing measures that attempt to elicit individual priors \citep{charness2021experimental, garthwaite2005statistical, goldstein2014lay, johnson2010methods, schlag2015penny, stefan2020practical}. One approach elicits a single numerical value, range, or quantile estimate for the probability of a particular outcome \citep{manski2004measuring, o2019expert}. Another is to allow a `budget' of probability mass that can be assigned to a set of possible outcomes \citep{goldstein2008choosing, johnson2010valid}. Others capture belief over a range of possible outcomes using sliders to adjust histogram bar heights \citep{franke2016does}. Sliders can also be used to adjust parameters of a distributions over a continuous range of values \citep{jones2014prior}. However, many of these methods are either impoverished in the number of outcomes that can be represented or in the specificity allowed for any particular estimate. Having participants adjust limited parameters of a particular distribution requires an unwarranted assumption that beliefs are indeed represented by said distribution. Our method to elicit beliefs attempts to address these limitations by allowing participants to make estimates over a large number of potential outcomes while minimizing restrictions on the specificity of each estimate, without relying on any particular parameterization.

In our presented experiments, each event is portrayed as a ball drop landing in one of 40 slots. Participants beliefs are thus represented as histograms of 40 bars, where the relative heights of the bars indicate the participants' relative expectation of where future ball drops will land. These histograms are produced with a single click-and-drag of a computer mouse or a touchscreen over the histogram bins. This allows participants to easily express their updated beliefs at each trial of our task without being encumbered by a tedious and effortful response format. We elected to use ball drops as an intuitive narrative through which to present new data that would make participants update their beliefs. However, the presented events need not necessarily be ball drops. In principle, our presented method can be used for any situation where discrete events could be easily represented on a uni-dimensional ordered spatial domain over which a histogram could be drawn. This affords the opportunity to explore statistical learning in either a domain specific, or a domain general manner. We emphasize that the the most useful feature of our proposed methodology is the ability to capture individual participant priors in a rich yet theory-agnostic manner, and continuously monitor how they update as learning occurs. 

Here we present three experiments that demonstrate the utility drawing belief distributions to measure mental models and statistical learning. Our method can be used to cluster participants by their priors to predict learning outcomes (Experiment 1), measure how participants update to unannounced distribution changes (Experiment 2), and to measure the capacity to represent physically implausible probability distributions that are dynamically defined according to participant input (Experiment 3). We also explore whether the priors are reliable representations that can meaningfully characterize prior beliefs on an individual level, and do not simply reflect the intuitive physics of our task (Experiment 3).

\section{General Method}
\subsection{Task Environment}

We developed a computerized version of the game ``Plinko'', a modern incarnation of Galton’s Bean Machine \citep{galton1894natural} featured on the American game show \emph{The Price is Right}. In Plinko, balls fall through pegs to land in slots below. In our task, participants view a triangle of 29 black pegs drawn on the computer screen while a red ball is dropped from the top peg. The ball follows a cascading path to land in one of the 40 slots below the pegs. The algorithm used to determine ball drop trajectory varied by experiment.

\begin{figure}
  \includegraphics[width=\textwidth,height=\textheight,keepaspectratio]{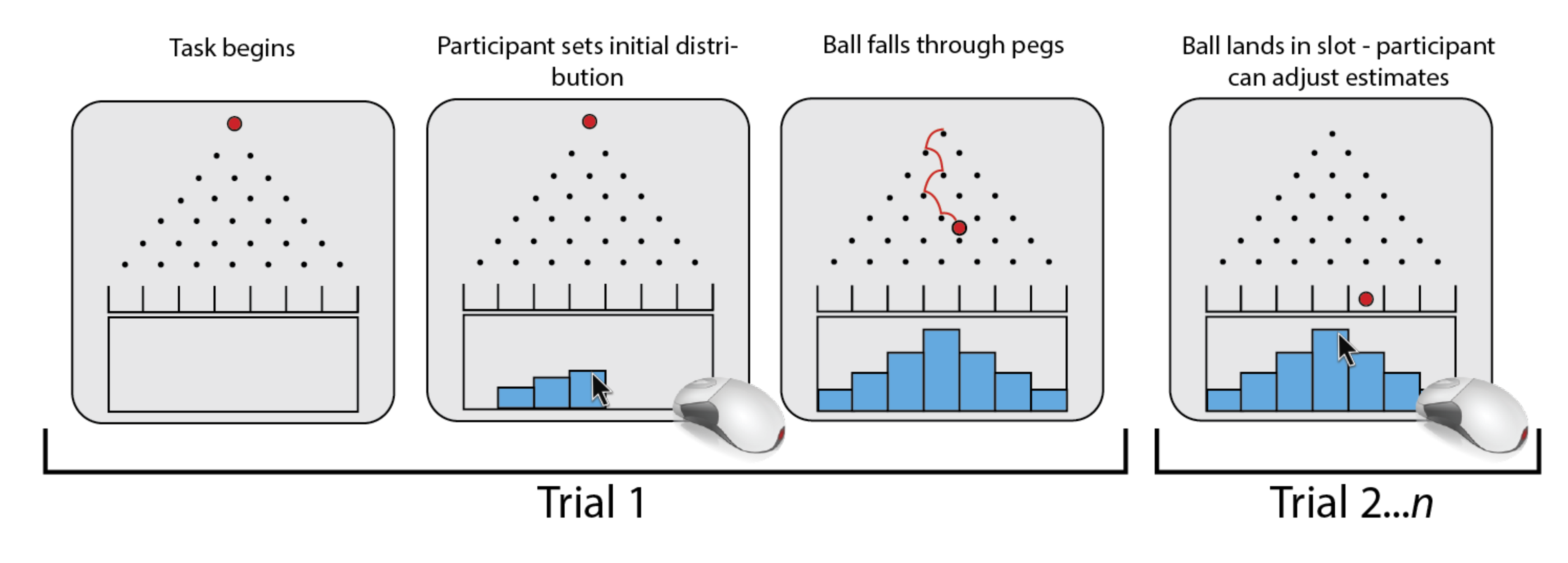}
  \caption{Schematic of the Plinko task. A red ball `fell' through a pyramid of to land in one of the 40 slots below. Participants first drew bars using the computer mouse (or their finger on a touch screen) to indicate the most likely locations the ball would land in – higher bars indicate an expectation of higher likelihood. Seven slots are pictured here for illustrative purposes; the number of slots and pegs can be adjusted in the task’s source code.}
  \label{fig:plinko}
\end{figure}

In one version, the participants provided their estimates of the ball drop distribution by clicking and dragging the mouse below the slots to draw a histogram. In another, participants used a touchscreen to draw histograms with their fingers. We told participants that higher bars represented a higher probability that a ball would fall in a slot, lower bars a lower probability, and that drawing no bar represented zero probability. Participants could draw bars under one, some, or all slots, provided at least one bar was drawn on the screen before proceeding to the next ball drop. The total area of participant drawn histograms was not restricted, so long as the bars fit within the available display (Figure \ref{fig:plinko}). We programmed the task in Python using the PsychoPy library \citep{peirce2009generating}. A demo and source code for the task is freely available online at our OSF repository, https://osf.io/dwkie.

\subsection{Participants}

We analyzed preexisting data that were collected across three separate experiments. All 335 participants across the three experiments were University of Waterloo undergraduates. The University of Waterloo’s Office of Research Ethics cleared all study protocols and participants gave informed consent before participating. Some participants in the dataset had missing or incomplete data. We only analyzed participants with complete trial data. Some participants had missing demographic data (noted in each experiment). We still analyzed participants with missing demographic data if they had complete trial data. All data and analyses are available at our OSF repository, https://osf.io/dwkie.

\subsection{Data Analysis}

We measured participant performance by computing the angular similarity between the representative Euclidean vectors \citep{georgopoulos1986, cer2018, angsim} of the participants' drawn histograms, and the known experimental ball drop distributions. This results in a similarity score that ranges from 0 when participant histograms share no mass with the experimental reference distribution, to 1 when participants' estimates are proportional to the experimental reference distribution (see Appendix). There exist many methods to measure ``distance'' between probability distributions, including Kullback-Leibler Divergence \citep{kullback1951information}, Bhattacharyya Distance \citep{bhattacharyya1943measure}, and Earth Mover Distance \citep{rubner2000earth}, to name only a few. However, there is no clear ``correct'' measure that one should use to compare distributions. Moreover, most approaches require comparisons between normalized probability distributions. Our goal in this paper is to make as few assumptions about our data as possible. As such, all analysis is performed using the raw, unnormalized bar heights from participant drawn histograms wherever possible. We elected to use angular similarity because it does not require any data prepossessing (normalization) and because of its overall simplicity (see Appendix). All aggregate learning curves are fitted local regression curves ($\alpha = 0.1$), a built-in method from the ggplot2 package \citep{ggplot}.

We performed all analyses in R \citep{r}, using the data.table \citep{dt}, magrittr \citep{mag}, pracma \citep{prac}, plyr \citep{ply}, ggplot2 \citep{ggplot}, ggpubr \citep{ggpub}, rstatix \citep{rstx}, gridExtra \citep{gridx}, gridGraphics \citep{gridgraph}, purrr \citep{pur}, GmAMisc \citep{gma}, dynamicTreeCut \citep{langfelder2008}, Matrix \citep{mat},  emdist \citep{emd}, and knitr \citep{knit} packages.

\section{Experiment 1: Clustering priors}

In this experiment, we explore the structure of prior variability by demonstrating three methods of clustering priors. The prior beliefs we hold impact how we interpret future events \citep{green2010, hock2005, lee2013, patrick2014, bianchi2020, stottinger2014priors}. Consequently, some patterns of decision making are necessarily ``better'' than others given a particular task. We therefore also consider the influence a prior may have on statistical learning by comparing learning accuracy across clusters of priors. We also consider the degree of mental model smoothing humans employ when integrating new statistical evidence. That is, do participant ball drop estimates approach the literal histogram of presented ball drops, or do they instead approach a smooth idealized distribution that `summarizes' the discretely presented stimuli comparable to perceptual averaging \citep{ariely2001, corbett2011, albrecht2012}? 

\subsection{Method}

We analyzed the data of 266 University of Waterloo undergraduates (3 missing demographic data, 197 female, mean age = 19.96, $SD$ = 2.30 years) to measure the predictive value of three prior clustering methods on learning accuracy. 

Participants performed a series of tasks as part of a larger study investigating exploratory behaviour as a function of boredom proneness. These included two versions of a virtual berry picking foraging task \citep{struk2019}, a ‘connect-the-dots’ problem solving task, and a version of a word search task intended to function as a cognitive ‘foraging’ task in which participants searched an array of letters to make words. Participants also completed a version of the Plinko task. While the berry picking, connect-the-dots, and word search tasks were counterbalanced in order, the Plinko task was always performed last. Each task took around 10 minutes to complete. The series of tasks were completed on a touchscreen placed on a flat table, and inclined at approximately 25 degrees. Some tasks in this larger study examined the relationship between foraging patterns and gene expressions, motivating a larger sample size in this experiment than that of Experiments 2 and 3. 

We asked participants to provide their estimate of the ball drop distribution \emph{before} seeing any ball drops. Following collection of the one initial prior, we asked ``How confident are you that your bars reflect the likelihood that a ball will fall in any of the slots?'' We recorded confidence with a sliding bar from ``Not Confident'' to ``Very Confident'', translating to a confidence score ranging from 0 to 1 (inclusive), where 1 is most confident. One participant did not have accompanying confidence data and was thus omitted from analysis of confidence data.

The task continued for 50 trials. Each trial consisted of a single ball drop, and participants could modify their estimate as they saw new events. Participants were not informed that there was any particular structure to the distribution of ball drops. Each participant observed an identical sequence of ball drops, regardless of their reported prior or trial-by-trial predictions. The sequence of ball drops followed a unimodal distribution centered over the 18th slot with a standard deviation of 4.84 slots. We elected to present an identical and representative ball drop sequence across participants to reduce noise. We are interested in comparing performance across individual differences in priors, so reducing possible effects of idiosyncratic ball drop sequences is particularly important here.

We considered two possible candidate ``reference'' distributions for this analysis. First, the histogram of the literal ball drop sequence given to all participants in this experiment. Second, a normal distribution with the same mean and standard deviation (stated above) as the observed sequence of ball drops (Figure \ref{fig:smoothOrRealPlotsCombo}A). To compare candidate reference distributions, we plotted aggregate learning curves, and compared final trial learning accuracy with respect to each candidate reference distribution.

We applied three distinct methods for clustering participant priors to explore the relationship between priors and learning accuracy. We adopted the first method from \citet{shu2011} to cluster priors according to their general shapes. Originally created for 2D shape matching and image retrieval, we omitted the initial steps of the algorithm that converts a binary shape image into a contour of points distribution histogram (CPDH) \citep{shu2011}. We instead used the 40 unnormalized bar heights from each prior to produce each CPDH. All other steps were identical. This method clusters priors based on general shape, and is insensitive to changes in scale, translation, and orientation, which was important in the context of our goal. That is, two participants may represent a prior in the shape of a Gaussian distribution, but do so with different bar heights. Under this method, these two participants would be considered as members of the same cluster, which would not necessarily be the case with other clustering algorithms. This method operates by producing a dissimilarity matrix between each CPDH using Earth Mover's Distance (EMD) \citep{shu2011}. Note that EMD is applied to pairwise combinations of the CPDHs produced by the algorithm, not the raw histogram priors drawn by the participants. We then performed a hierarchical cluster analysis on the resulting set of pairwise participant prior dissimilarities. We created a dendrogram using the \emph{hclust} built-in R function, and cut the dendrogram branches to define our clusters using the \emph{cutreedynamictree} R function from the Dynamic Tree Cut package \citep{langfelder2008}. Originally designed for detecting clusters in genomic data, the \emph{cutreedynamictree} R function uses an iterative process to combine and decompose clusters within a dendrogram until the number of clusters stabilizes \citep{langfelder2008}. We selected function parameters to ensure that all priors were assigned to a cluster.

The second clustering method was a manual classification performed by an author. Each prior was visually classified as either a ``Gaussian'', ``Bimodal'', ``Uniform'',  ``Jagged'', ``Skewed'', or ``Trimodal''. This method was used to explore our own subjective intuitions about the patterns we saw in the data. We did not consult participant performance to form these clusters -- only the shape of each participants' prior. For both the CPDH and visual clusters, we plot aggregate learning curves and compare the final trial learning accuracy across prior clusters.

Our third clustering method categorized participants in a reverse manner to the first two methods. Here, we clustered participants by their \emph{final trial learning accuracy}. This requires a method that clusters participants on the basis of a single numerical value (final trial accuracy), rather than a hand drawn ball drop estimate (participant priors) which was required by our first two clustering methods. We elected to use the Jenks' natural breaks method implemented by the GmAMisc R package \citep{gma} for this purpose. We then visually explored differences in participant priors across the learning accuracy clusters.

\subsection{Results and Discussion}

\subsubsection{Participant estimates are idealized and smooth, not literal representations of presented data}

We performed a two-way repeated measures ANOVA comparing participant ball drop predictions to each candidate reference distribution at the first and final trials. We found a main effect of reference distribution $F(1,265) = 1556.49, p  < .001$, a main effect of trial $F(1,265) = 190.91, p  < .001$, and an interaction between reference distribution and trial $F(1, 265) = 62.55, p  < .001$. Participants exhibit a learning effect when measured against each candidate reference distribution. Participants' final trial estimates ($M =  0.55$, $SD = 0.09$) were more similar to the histogram of observed ball drops than their first trial estimates ($M =  0.42$, $SD = 0.15$),  $t(265) = 14.10, p  < .001$. This is also true when using a smooth normal distribution with the same mean and standard deviation as the histogram of observed ball drops, where final trial similarity ($M =  0.67$, $SD = 0.12$) is greater than first trial similarity ($M =  0.50$, $SD = 0.19$), $t(265) = 13.45,  p  < .001$

The interaction between reference distribution and trial implies that the greater similarity to the smooth reference distribution increases over time. The additional accuracy to the smooth reference distribution is not constant. This suggests that participants approach the idealized distribution faster by smoothing trial-by-trial ball drop data and incorporating new evidence into a simplified representative model rather than accumulating literal and discrete events. It is also possible that the higher similarity to the idealized distribution is a result of drawing ball drop estimates by dragging a computer mouse or a finger along a touchscreen; a smooth continuous curve is easier to draw than jagged histogram. However this ease-of-drawing argument does not account for the relative increase in prediction accuracy in later trials for the idealized distribution. It may not be surprising that participants reproduce an idealized distribution instead of a literal accumulation of individual events, but this result justifies our intuition that we should compare participant data to the distribution that generates our events and not the events themselves.

\begin{knitrout}
\definecolor{shadecolor}{rgb}{0.969, 0.969, 0.969}\color{fgcolor}\begin{figure}
\includegraphics[width=\textwidth,height=0.87\textheight,keepaspectratio]{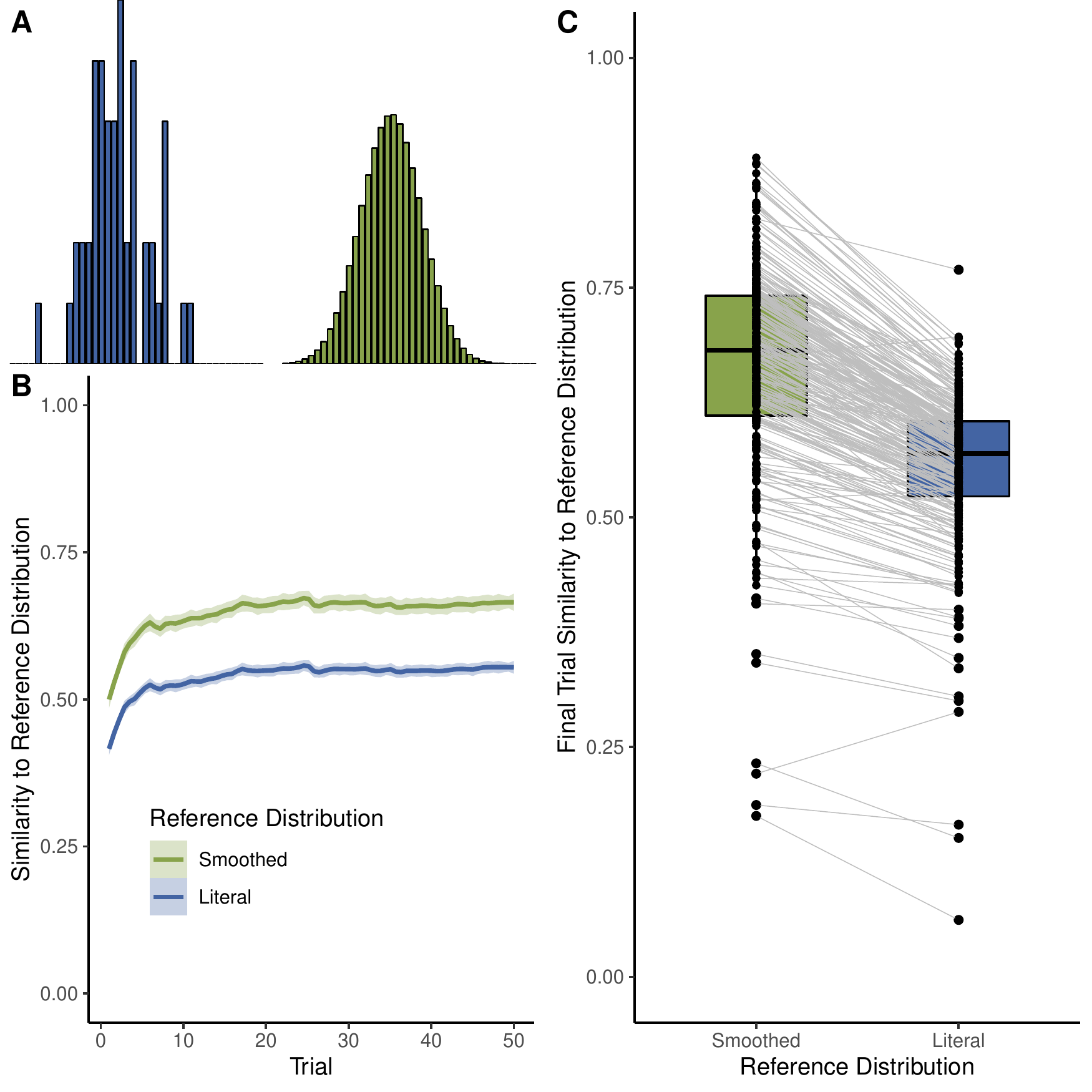} \caption[A]{A: A histogram of the 50 presented ball drops (blue) and a normal distribution with the same mean and s.d. as the literal ball drops (green). B: Average participant learning curve with respect to the smoothed normal distribution (green) and the literal ball drop histogram (blue), +/- 95$\%$ CI. C: Similarity at final trial to the smoothed normal distribution (green) and the literal ball drop histogram (blue).}\label{fig:smoothOrRealPlotsCombo}
\end{figure}

\end{knitrout}

\subsubsection{Priors can be clustered by shape matching algorithms to disambiguate learning accuracy}

Our exploratory hierarchical cluster analysis produced three distinct categories of participant priors: concave unimodal ($n = 97$), bimodal ($n = 86$), and convex unimodal ($n = 83$) (Figure \ref{fig:allEMDFig}A). All participant priors were assigned to a cluster. Note that these clusters formed organically -- we did not assume theme to exist, or set the algorithm to produce clusters that resemble well-known distribution families. The names we have assigned to the resulting clusters are for descriptive purposes only.

We performed a one-way ANOVA comparing participants' final trial learning accuracy to the idealized reference distribution, grouped by CPDH prior cluster. Cluster membership did not indicate a statistically significant difference in final trial accuracy, $F(2, 263) = 2.69, p  = .070$. However, the concave-unimodal cluster presents a visual and numerical separation in learning accuracy from the other two clusters (Figure \ref{fig:allEMDFig}B). Self-reported confidence in priors did not vary between CPDH prior clusters $F(2, 262) = 0.85, p  = .427$.

Our choice of this shape matching algorithm reflects our belief that the distributional family matters most for finding patterns in a population of individual priors, not any particular parameter value of central tendency. This belief is consistent with other work \citep{griffiths2006optimal}. We believe that this cluster analysis should be performed on the unnormalized, not normalized,  bar heights of participant priors  because the CPDH algorithm clusters by shape. Normalizing priors maintains the relative size of each bar height, but alters the visual appearance of the overall shape. By using the raw bar heights produced by participants, we minimize distortions and maintain the integrity of this particular shape-based method.

\begin{knitrout}
\definecolor{shadecolor}{rgb}{0.969, 0.969, 0.969}\color{fgcolor}\begin{figure}
\includegraphics[width=\textwidth,height=0.87\textheight,keepaspectratio]{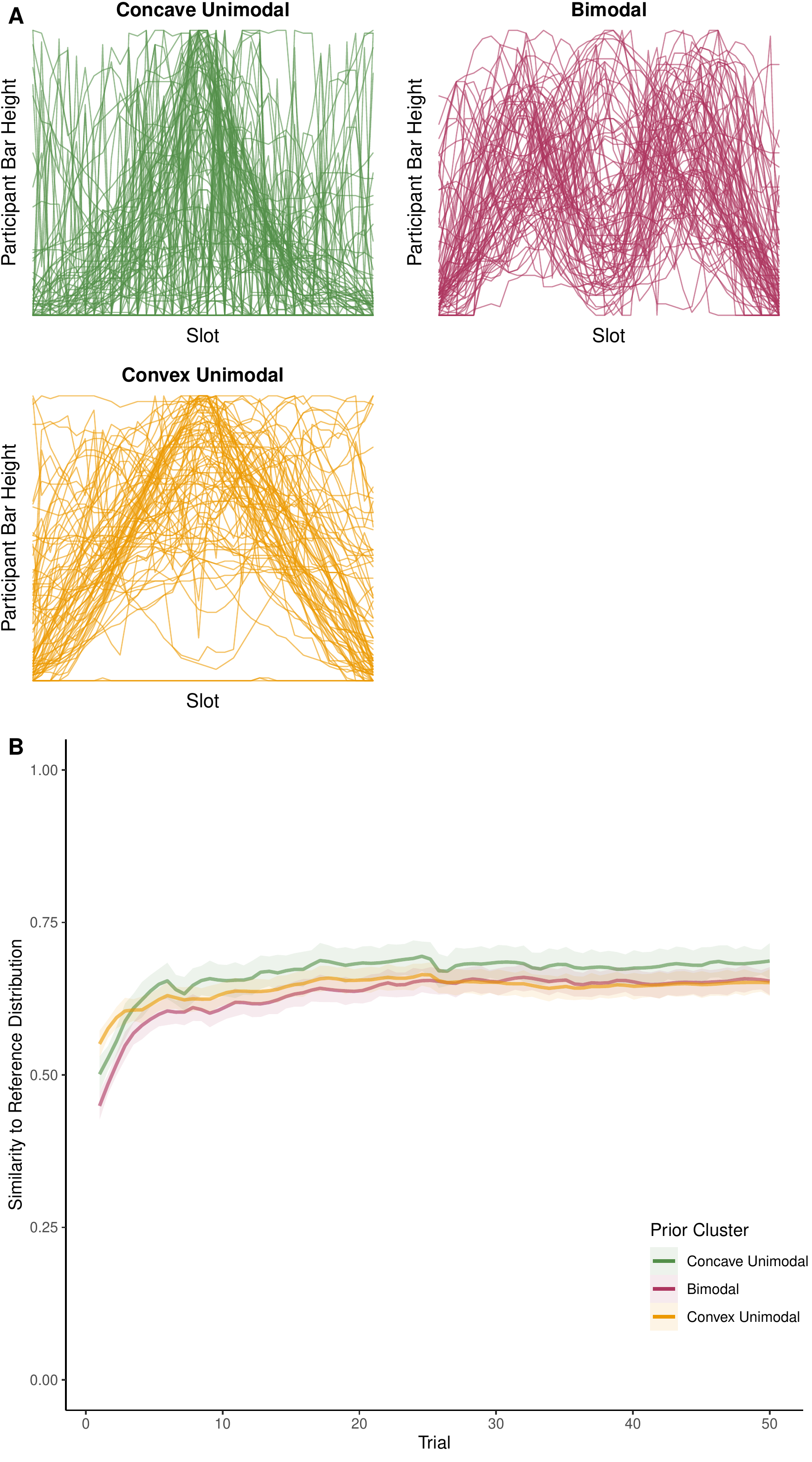} \caption[A]{A: Line plots of individual priors are presented together, grouped by cluster. Hierarchical CPDH clustering results in three clusters of participant priors: concave unimodal, bimodal, and convex unimodal. B: Aggregate learning curves, split by CPDH clusters, +/- 95$\%$ CI.}\label{fig:allEMDFig}
\end{figure}

\end{knitrout}

\subsubsection{Manually clustered priors disambiguate learning accuracy}

Our subjective post hoc clustering of participant priors yielded 6 unique clusters: ``Gaussian'' ($n = 123$), ``Bimodal'' ($n = 86$), ``Uniform'' ($n = 31$), ``Jagged'' ($n = 9$), ``Skewed'' ($n = 8$), and ``Trimodal'' ($n = 6$). For our analyses, we collapsed the ``Jagged'', ``Skewed'', and ``Trimodal'' into an ``Other'' cluster ($n = 23$), resulting in 4 final clusters (Figure \ref{fig:allAlexPlots}A). Of the total 266 participants in this study, 3 did not have accompanying manual cluster assignments, and were thus omitted from this analysis.

We performed a one-way ANOVA comparing participants' final trial similarity to the idealized reference distribution, grouped by manual cluster. Final trial learning accuracy varied across clusters, $F(3, 259) = 5.81, p  < .001$ (Figure \ref{fig:allAlexPlots}B). Post hoc comparisons using the Tukey HSD test indicated that the mean final trial similarity for the ``Gaussian'' prior group $(M = 0.69, SD = 0.10)$ was greater than the ``Uniform'' prior group $(M = 0.60, SD = 0.15)$, $p  < .001$. No other pairwise comparisons were significant, $ps \geq 0.10$.

Self-reported confidence varied between manual prior clusters $F(3, 258) = 3.53, p  = .015$. Post hoc comparisons using the Tukey HSD test indicated that the mean confidence rating for the ``Gaussian'' prior group $(M = 0.56, SD = 0.23)$ was greater than that of the ``Other'' group $(M = 0.41, SD = 0.15)$, $p  = .018$. No other pairwise comparisons were significant $ps \geq 0.27$

\begin{knitrout}
\definecolor{shadecolor}{rgb}{0.969, 0.969, 0.969}\color{fgcolor}\begin{figure}
\includegraphics[width=\textwidth,height=0.87\textheight,keepaspectratio]{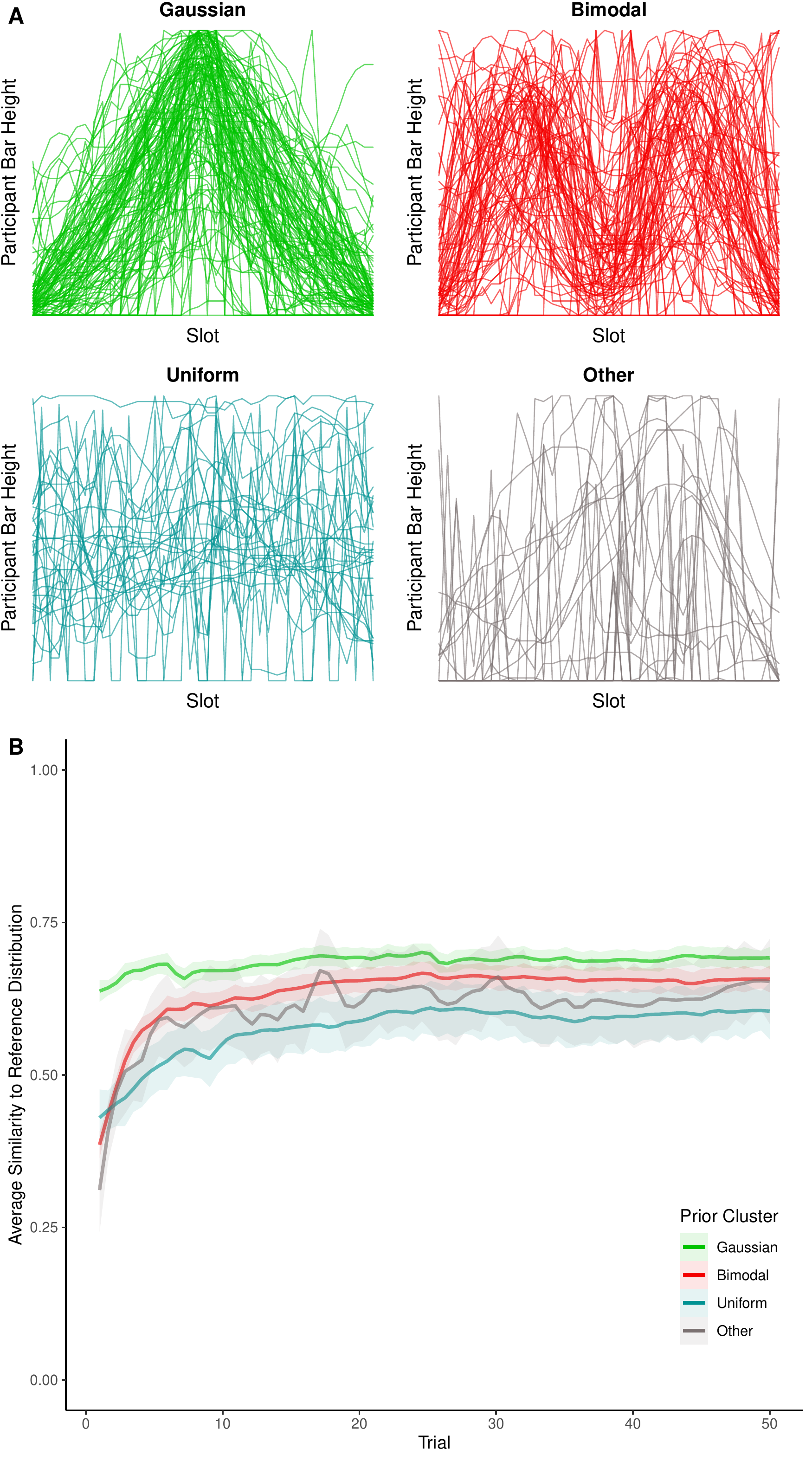} \caption[A]{A: Line plots of individual priors are presented together, grouped by manual clusters:  Gaussian, bimodal, uniform, and other. B: Aggregate learning curves, split by manual clusters, +/- 95$\%$ CI.}\label{fig:allAlexPlots}
\end{figure}

\end{knitrout}

It may be appropriate to treat participants in the ``Other'' ($n = 23$) prior cluster as outliers. We repeated the above analysis after excluding participants in the ``Other'' prior cluster. Final trial learning accuracy varied across clusters, $F(2, 237) = 8.73, p  < .001$. 
Post hoc comparisons using the Tukey HSD test indicated that the learning accuracy for the ``Gaussian'' prior group $(M = 0.68, SD = 0.10)$ was greater than that of the ``Uniform'' group $(M = 0.58, SD = 0.14)$, $p  < .001$. Differences between the ``Gaussian'' and ``Bimodal'' clusters $(M = 0.63, SD = 0.12)$, $p  = .061$, and the ``Uniform'' and ``Bimodal'' clusters, $p  = .055$, were not significant, though visually apparent (Figure \ref{fig:allAlexPlots}B). Self-reported confidence between manual prior clusters (excluding ``Other'') showed no differences $F(2, 236) = 2.10, p  = .124$. 

\subsubsection{Distinct priors are indicated when participants are clustered by learning accuracy}

Unlike the clustering methods above, we reversed our approach and clustered participants by final trial learning accuracy, not by properties of their priors. We used Jenks' natural break method to group participants into three clusters on the basis of their final trial similarity (Figure \ref{fig:showReverseClusters}).  If features of participants' prior influences learning accuracy, then differences in learning accuracy may also indicate differences in priors. The worst performing cluster (n = 29), contained participants with final trial accuracy between 0.175 and 0.528. The middle performing cluster (n = 127) contained participants with final trial accuracy between 0.528 and 0.697. The best performing cluster (n = 110) contained participants with final trial accuracy between 0.697 and 0.891. The model's goodness of fit was 0.774, relative to a max goodness of fit of 0.999 reached with 75 clusters. 

\begin{knitrout}
\definecolor{shadecolor}{rgb}{0.969, 0.969, 0.969}\color{fgcolor}\begin{figure}
\includegraphics[width=\textwidth,height=0.87\textheight,keepaspectratio]{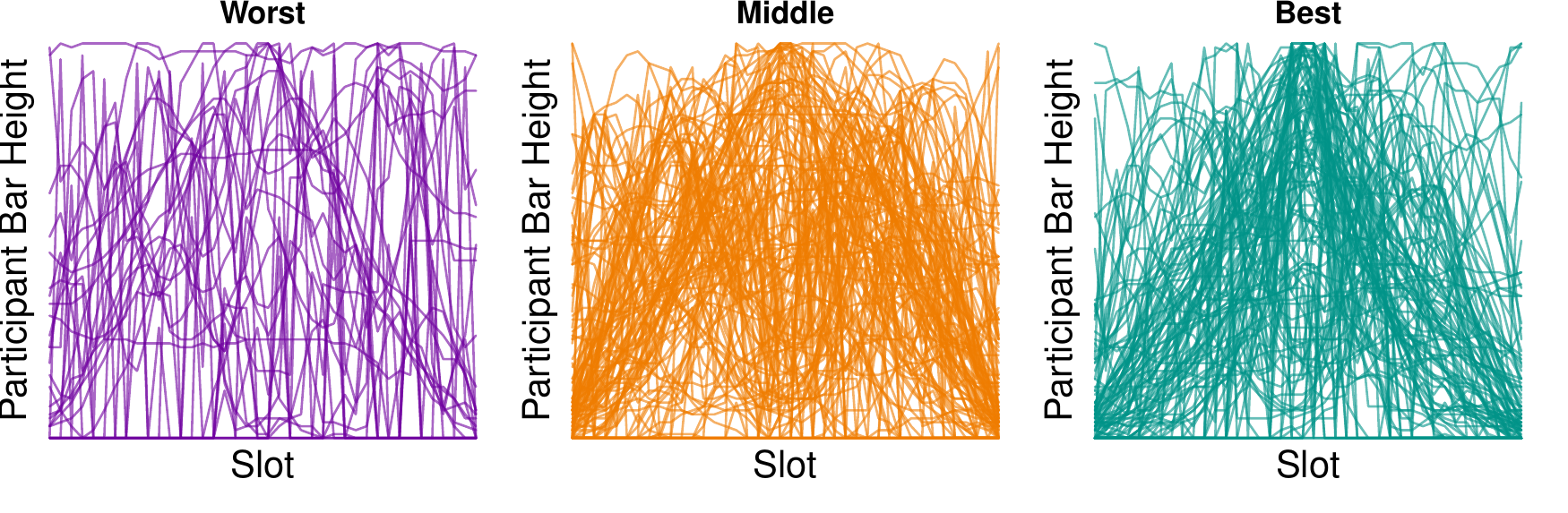} \caption[Individual Participant priors clustered by final trial accuracy]{Individual Participant priors clustered by final trial accuracy. Clustering participants by performance indicates cluster-specific regularities: worst performing group has irregular priors, middle performing group has unimodal or bimodal priors, best performing group has mostly concave unimodal priors.}\label{fig:showReverseClusters}
\end{figure}

\end{knitrout}

By visual inspection, each reverse cluster presents unique features. The first cluster of the worst performers presents fewer regularities than the two better performing clusters. The middle performing group contains both unimodal and bimodal priors, while the best performing group mostly contains concave unimodal priors. Self-reported confidence between reverse prior clusters showed no differences $F(2, 262) = 1.02, p  = .363$.

The three presented clustering methods demonstrate the potential for future research using Plinko as a testing methodology. We have demonstrated priors vary across individuals and that some set features of a participant's prior may indicate a participant's success in learning a presented probability distribution. Though it is still unclear what features of a prior may be of interest. The convexity of participant drawn curves is an important feature for image recognition and categorization, as demonstrated by our CPDH clustering results, but is a feature that may not appear relevant to a human rater. We encourage future research to consider adopting other computational methodologies from other disciplines to categorize participant priors, in addition to human rating. These results also demonstrate the importance of collecting rich histogram representations of priors. Subtle, but potentially relevant properties of individual priors may not be detected with coarser measures that restrict participants' ability to easily express their priors.

We refrain from declaring the `correct' set of families with which to separate individual priors. Here, we intend to demonstrate the wide array of methodologies that may be used to search for the correct set of prior clusters. There is difficulty in addressing concerns of convergent validity in the early stages of this work. That is, there is no clear mapping between the clusters found using each of our three methods since we have no clear prescriptive expectations as to what clusters ought to exist in a population of priors for a given task. Nonetheless, our results demonstrate that despite our emphasis on the importance of individual differences, regularities still likely exist in any given population of priors. Future work should address the reliability of the method that best identifies these regularities.

Our presented data is also fundamentally limited in its ability to express learning accuracy across prior clusters. We do not know whether the differences in learning outcomes are due to trait or state differences. It may be that participants who hold certain priors are better learners, or it may be that participants with priors that are more similar to the forthcoming distribution are better able to learn that distribution. However, this potential effect is masked in our current experiment, because those who may be the best learners of a particular distribution also have priors that are the most similar to that distribution. We therefore elected to not analyze learning rate here, as a low learning rate may only indicate a `better' prior, rather than poorer learning. Instead, we propose a future experiment that uses each participants' prior to define the distribution each participant is to learn. In this experiment, each participant is presented with a unique distribution, but one that is a consistent, fixed distance from their original estimate. Such an experiment may necessitate an initial assumption as to how humans measure similarity between probability distributions, but would also increase the validity of analyses of learning rate and learning accuracy.

\section{Experiment 2: Updating to changes}

In Experiment 1 we explored how priors may predict the learning ability of participants. In Experiment 2, we investigate participants' ability to update their mental model of a learned probability distribution, and how experimental environments influence this ability. Previous research indicates that it is a non-trivial problem to determine \emph{when} and to \emph{what extent} a mental model has been updated \citep{oreilly2013}. In this Experiment, we detect model updates by measuring participants' trial-by-trial accuracy, and median ball drop estimates when learning multiple probability distributions presented in sequence. We then compared how updates to ball drop predictions are influenced by the presence of a break from the task before the ball drop distribution changes.

\subsection{Method}

We tested 39 University of Waterloo undergraduates (2 missing demographic data, 21 female, mean age = 20.07, $SD$ = 2.06 years) to measure how participants update to changes in the presented ball drop distribution. Due to an error in data collection, raw participant bar heights were missing for 24 of the total 39 participants. We therefore performed all analysis for this experiment using normalized participant estimates rather than unprocessed participant estimates. Since we used the angular similarity between the representative Euclidean vectors of participant estimates and the appropriate reference distribution to define learning accuracy, our analysis is invariant to the scale of our data (see Appendix). Therefore, using the normalized bar heights instead of the unprocessed bar heights of participant estimates had no impact on our results. This may not be the case for future users of our data if they elect to use an alternative measure.

We presented participants with four sequences of 100 ball drops (400 trials in total). Each sequence of ball drops was generated from a distinct probability distribution: 1) a wide normal distribution ($M$ = slot 17, $SD$ = 6 slots), 2) a narrow normal distribution ($M$ = 30, $SD$ = 2), 3) a bimodal distribution made of an equal mix of two normal distributions ($M$ = 9, $SD$ = 3) and ($M$ = 27, $SD$ = 3), and 4) a positively skewed (Weibull) distribution ($\alpha = 6, \beta = 1$) (Figure \ref{fig:allNormFig}A). Each participant was exposed to the exact same sequence of ball drops. We elected to present an identical and representative ball drop sequence across participants to reduce noise. We are interested in comparing how participants update to new distribution across break conditions, so reducing possible effects of idiosyncratic ball drop sequences is particularly important here for similar reasons to Experiment 1.

We assigned participants to one of two conditions. Participants in the ``break'' condition ($n = 20$) were given a break between each ball drop distribution, but were not told the significance of the break; that is that it signaled a change in the ball drop distribution. Participants pressed the space bar to continue the task. Participants in the ``continuous'' condition ($n = 19$) observed an identical sequence of ball drops to the ``break'' group, but were given no breaks between ball drop distributions.

To determine how effectively each group adjusted to each new ball drop distribution, we compared the average similarity of participant estimates at the final trial of each ball drop sequence to the relevant ball drop distribution. We also visually inspected heat maps indicating the median normalized bar heights at each trial.

\subsection{Results and Discussion}

We performed a two-way mixed ANOVA to determine whether participants' ball drop estimate accuracy at the final trial of each 100 trial distribution varied across break conditions. We found a main effect of break condition, $F(1,37) = 5.73, p  = .022$  and ball drop distribution, $F(3, 111) = 33.88, p  < .001$, and an interaction between the two, $F(3, 111) = 7.38, p  < .001$ (Figure \ref{fig:allNormFig}B). Pairwise t-tests between break conditions at each ball drop distribution revealed that participants who were given a break between ball drop distributions predicted the narrow unimodal ($p  = .004$) and Weibull ($p  = .005$) distributions better than participants who got no breaks. There were no group differences in accuracy for wide unimodal ($p  = .484$) and bimodal ($p  = .455$) ball drop distributions.

Figures \ref{fig:allNormFig}C and \ref{fig:allNormFig}D demonstrate a ``hangover'' or ``hysteresis'' effect \citep{hock2005} in the continuous condition, where participant ball drop estimates of previous distributions are integrated into the next distribution. The effect is not seen in the break condition, where participants appear to treat each sequence of ball drops (separated by a break) independently. No differences in learning accuracy between the break conditions should be expected in the wide unimodal ball drop sequence, since participant experience was identical for both groups until the first distribution change. The observed ``hangover'' effect may also increase learning accuracy in some cases. A new ball drop distribution that is more similar to the aggregate pattern of all previous ball drops than a participant's prior will reduce the benefit of ``starting fresh'' when facing a new distribution of ball drops. This may explain the lack of any difference in learning accuracy between break conditions for the third presented (bimodal) distribution.

The results from this experiment demonstrate that participants can effectively learn multiple probability distributions presented in sequence. We have also demonstrated that the ability to update previously established beliefs can be manipulated by experimental features other than just the presented ball drop distributions. Future work using this task could explore the role priors play in updating to new ball drop distributions. For example, Plinko provides a convenient mechanism with which to test the true ``unbiased'' nature of a uniform prior: are participants beginning with a uniform prior (or pushed to a uniform estimate) better able to detect changes in ball drop distributions?

\begin{knitrout}
\definecolor{shadecolor}{rgb}{0.969, 0.969, 0.969}\color{fgcolor}\begin{figure}
\includegraphics[width=\textwidth,height=0.87\textheight,keepaspectratio]{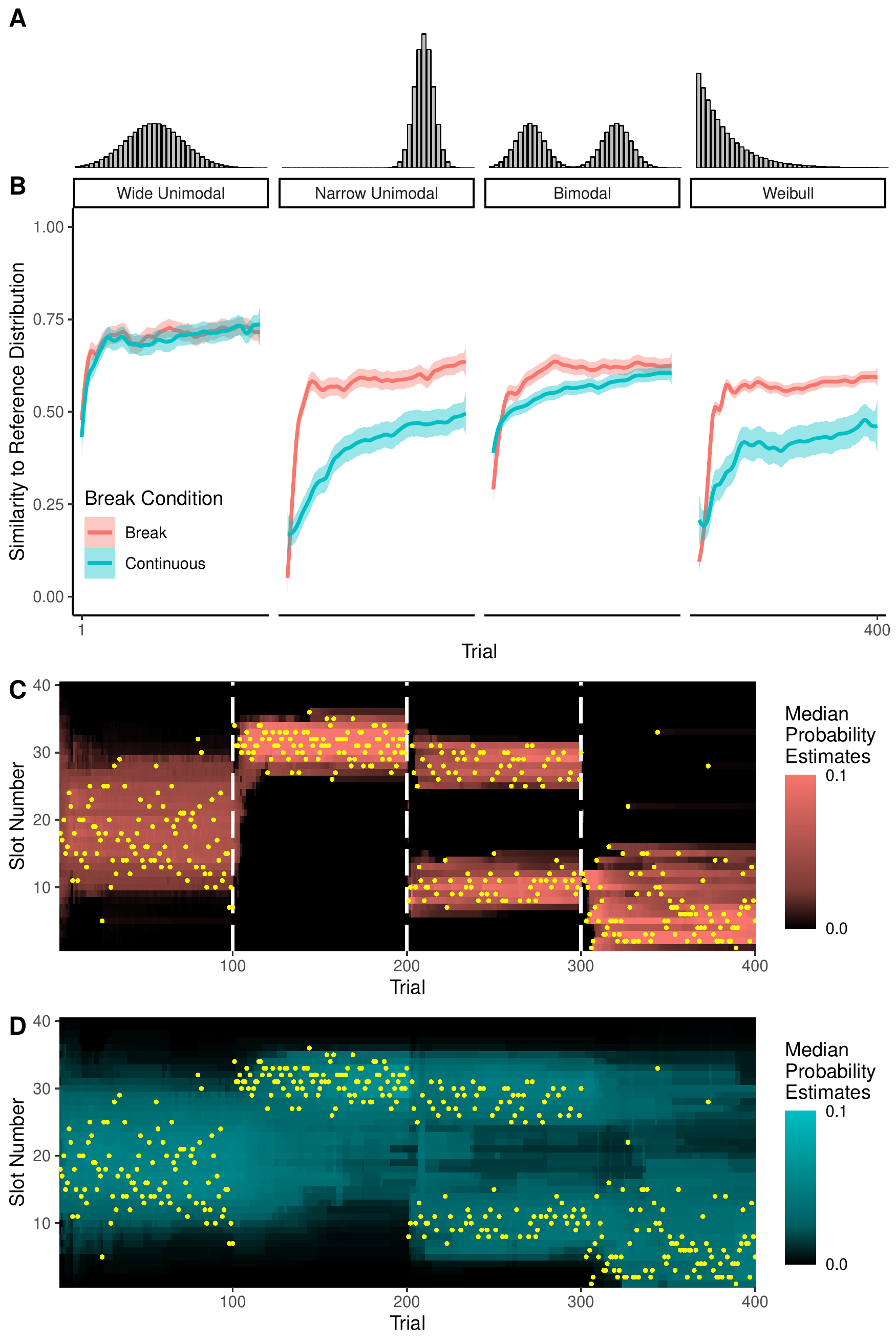} \caption[A]{A: Participants were presented with a sequence four ball drop distributions of 100 ball drops each. B: Aggregate learning curves across the four ball drop distributions for each break condition group, +/- 95$\%$ CI. Participants who were given a break between ball drop distributions had greater learning accuracy for the Narrow Unimodal, and Weibull ball drop disitributions. C: Heatmap of median normalized ball drop estimates for participants in the break condition. Each distribution is treated independently from the last. D: Heatmap of median normalized ball drop estimates for participants in the continuous condition. Unlike in C, participants exhibit hysteresis across ball drop distributions. Yellow dots represent the ball drop presented at each trial. The sequence of ball drops was identical accross participants and conditions.}\label{fig:allNormFig}
\end{figure}

\end{knitrout}

\section{Experiment 3: The reliability of prior belief measurements}

In Experiment 1, we showed that participant priors vary across individuals, but cluster around prototypical probability distributions. Further, participant success in a statistical learning task may be indicated by properties of their priors, justifying the importance of measuring priors at a high fidelity while making as few assumptions as possible. Experiment 2 demonstrated that our task can measure mental model updating by changing the presented ball drop distribution, and that task environments can influence participants' ability to update. In this experiment, we investigate whether the priors collected by Plinko are stable within each individual and limited to describing task-specific statistical learning.

A common approach to modeling probabilistic decision-making is to assume that all participants start a task with a homogeneous prior. Participant priors are often characterized either as representing maximal uncertainty (i.e., a uniform distribution; \citep{harrison2006encoding, mars2008trial, strange2005information}) or as approximating the process being estimated \citep{griffiths2006optimal, nassar2010approximately}. Attempts at empirically deriving priors \citep{gershman2016, spektor2018} still require assertions about the appropriate inferential model with which to estimate participant priors from participant data.  The results of Experiment 1 show that priors differ across individuals, and that these differences may play an important role in how participants learn new data.

Nevertheless, there is concern that the different priors observed in Experiment 1 are not stable and instead represent noise within this experimental context. We addressed the question of prior stability in this experiment by having participants provide two separate estimates of their prior (using deception -- see below). There is also concern that our task, which presents environmental contingencies with a physically plausible `correct' estimation (a binomial distribution), may be principally about itself. That is, asking participants to estimate distributions of balls dropping though pegs may only inform us of how statistical learning operates in this particular task. We address this concern here by presenting participants with a ball drop distribution that changes every trial, and is defined by their previous trial estimate. This produces a task environment that is markedly different than what would be expected in a real physical game of Plinko. We also explicitly ask participants to estimate the ball drop distribution of a physical real-life Plinko game and compare this estimate to the physically plausible binomial distribution.

\subsection{Method}

We tested 30 University of Waterloo undergraduates (20 female, mean age = 20.30, $SD$ = 3.22 years) to measure participant-reported prior beliefs of probabilistic estimates. To explore prior reliability, we \emph{twice} asked participants to provide their prior estimate of the ball drop distribution \emph{before} seeing any ball drops. Each prior was separated by a staged computer malfunction: After entering their first prior histogram and pressing the button to advance to the first trial, the screen went blank. The investigator told the participant that this was a common problem they knew how to fix. They then asked the participant to do a secondary task while they fixed the problem. The participant moved to a second computer and read a series of pronounceable non-words, each presented for one second. A large USB microphone with a glowing read LED was in front of the computer to enhance the deception. The distraction task lasted approximately two minutes. Upon completion of this deception, the participant returned to the first computer where the screen resembled the start of the experiment. The participant responded again with a ``first'' (now second) prior estimate before continuing with the rest of the task. Participants were debriefed at the end as to the nature of the deception, the reason for its inclusion, and were given the opportunity to rescind their permission to use their Plinko data. No participants rescinded permission. 

Following collection of both priors, we asked ``How confident are you that your bars reflect the likelihood that a ball will fall in any of the slots?''. We recorded confidence with a sliding scale from ``Not Confident'' to ``Very Confident'', translating to a confidence score ranging from 0 to 1 (inclusive), where 1 is most confident. The task continued for 99 trials with no further false malfunctions. Each trial consisted of a single ball drop, and participants could, but were not required to,  modify their estimates as they saw new events. Participants were not informed that there was any particular structure to the distribution of ball drops. Each ball drop was drawn from a distribution `opposite' to the participant's most recent distribution estimation. In this experiment, we coded the `opposite' discrete probability distribution by subtracting the participant bar heights from 100 (the maximum slot height) for each of the 40 slots. This opposite histogram was then scaled to contain a total area of 100 units to be a valid probability distribution from which to draw future ball drops. While the distribution ball drops are drawn from differ from the physically plausible distribution expected in a real-life game of Plinko, no individual ball drop violates the laws of physics as it cascades through the array of virtual pegs.

After completing all 99 trials, we had participants draw a distribution, as they had for the previous experimental trials, that represented the distribution of ball drops \emph{if this were a real physical game of Plinko} with a solid ball and pegs. We compared participant responses to the physically plausible binomial distribution, where the probability of a ball landing in the kth of 40 slots is ${39 \choose k} 0.5^k$.

To examine the reliability of participant priors, we first established the average similarity between each participant's prediction of the ball drop distributions recorded before and after the distraction. We then compared this average similarity to the means of 1000 random pairings of pre- and post- distraction priors. We also performed a correlation analysis to compare the participants' subjective rating of confidence in their priors to the reliability of their reported priors. Finally, we considered the role that individuals' understanding of the real-life physics of the game might play in our results by comparing the similarity of participants' prediction of the physical Plinko ball drop distribution to what would be expected in a physical game.

The uniform distribution becomes of particular interest under this conceptualization of `opposite'. It acts as an equilibrium, since the opposite of the uniform is the uniform. If a participant predicts a uniform distribution, the following ball drop will be drawn from a uniform distribution. If a participant predicts any non-uniform distribution, the following ball drop will be drawn from the opposite, but similarly non-uniform, distribution. Assuming a participant incorporates the current ball drop with previous ball drop data, a participant's prediction of future ball drops at trial $n+1$ will be more similar to the uniform than their prediction at trial $n$. We therefore set the uniform distribution as the benchmark or ``reference'' distribution when measuring learning accuracy for this experiment. We plot aggregate learning curves, comparing the similarity of the participant-drawn distribution to the reference uniform distribution at every trial. The aggregate learning curve is fitted with a Gompertz sigmoidal growth function \citep{gpz}. We elected to use a sigmoidal growth function instead of a standard exponential learning curve because sigmoidal fits allow for both a slowing in learning rate as maximal accuracy is reached, and for the maximal learning rate to occur at any point in time. The parameters of the Gompertz function, $data \sim Ae^{-e^{\mu e/A(\lambda - trial + 1)}}$, correspond to learning properties of interest. That is, $A$ defines the maximum similarity value reached, $\mu$ defines the maximum increase in similarity, and $\lambda$ defines the trial where $\mu$ occurs in the fitted model.

\subsection{Results and Discussion}

\subsubsection{Participants present reliable prior beliefs}

Figure \ref{fig:combinedPriorReliability}A compares the mean similarity (0.50) between first and second priors when we respect participant identity to a distribution of 1000 mean similarities of random permutations of first and second priors $(M = 0.33, SD = 0.02)$. By interpreting this distribution as a ``null'' distribution of chance similarity between priors, we can conclude that two priors from the same individual are more similar than two randomly selected priors, p $<$ 0.001. We thus conclude that 1) participants are heterogeneous in their priors, and that 2) participant reported priors are not merely `noise' as they represent something  persistent and unique to the individual.

\begin{knitrout}
\definecolor{shadecolor}{rgb}{0.969, 0.969, 0.969}\color{fgcolor}\begin{figure}
\includegraphics[width=\textwidth,height=0.87\textheight,keepaspectratio]{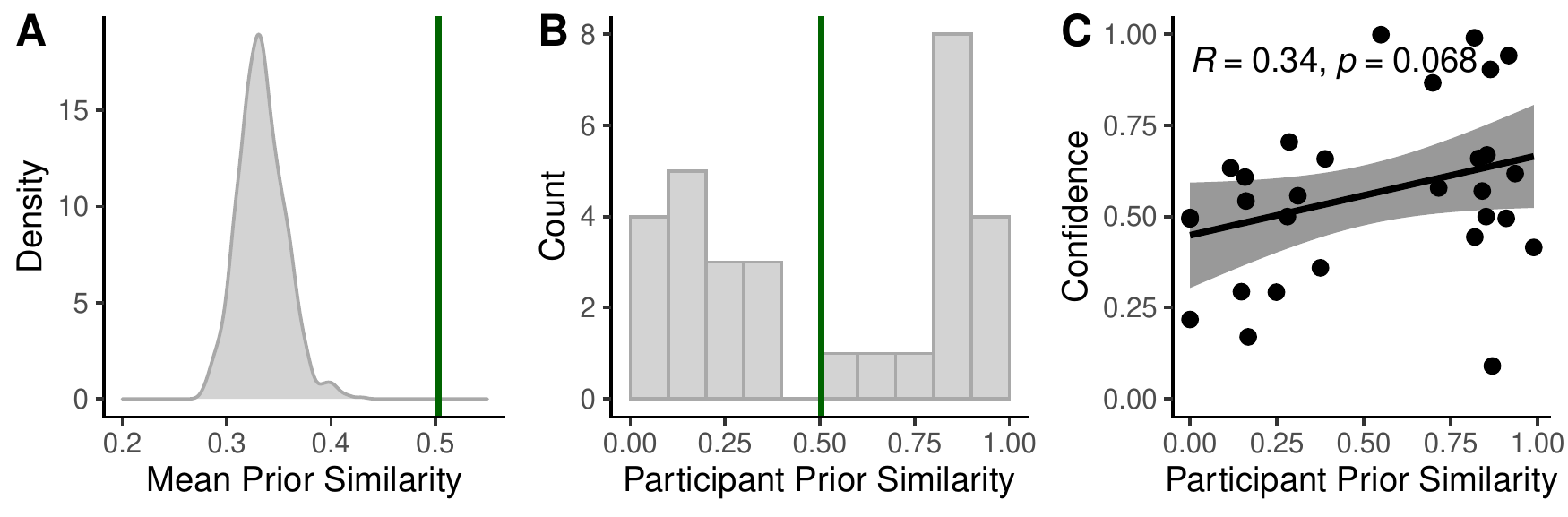} \caption[A]{A: The curve is the distribution of 1000 mean similarites of random permutations of first and second prior reports. The vertical line indicates the mean similarity for first and second priors when we respect participant identity. Similarity is higher when a participant's first reported prior is paired with their own second reported prior than with some other, randomly chosen, participant's second prior. This implies a prior has properties that are persistent over time, and unique to the individual. B: Histogram of paired prior similarites, respecting participant identity. The mean (vertical line) of this histogram is the vertical line of Panel A. The distribution of prior reliablity is bimodal. The lower mode is an artifact of our measure of similarity that is sensitive to lateral shifts of jagged distributions. C: There is a positive trend between prior reliability (similarity) and self-reported confidence in prior, +/- 95$\%$ CI.}\label{fig:combinedPriorReliability}
\end{figure}

\end{knitrout}

Participant prior reliability is bimodally distributed (Figure \ref{fig:combinedPriorReliability}B). Results of a Pearson's product-moment correlation indicates prior reproducibility may trend positively with confidence ratings, r(28) =  0.34, p  = .068 (Figure \ref{fig:combinedPriorReliability}C). Some participants have low measures of similarity between their first and second priors because their priors are jagged, with sparsely drawn slot estimates. Two jagged distributions are more prone to unusually low similarity values than two smooth distributions. Indeed, this appears to be the case. Participants with ``high'' (greater than 0.5) prior similarity exhibited smooth and similarly shaped priors (Figure \ref{fig:highLowSimAPPriors}A) whereas participants with ``low'' (less than 0.5) prior similarity made at least one jagged first or second prior (Figure \ref{fig:highLowSimAPPriors}B).

\begin{knitrout}
\definecolor{shadecolor}{rgb}{0.969, 0.969, 0.969}\color{fgcolor}\begin{figure}
\includegraphics[width=\textwidth,height=0.87\textheight,keepaspectratio]{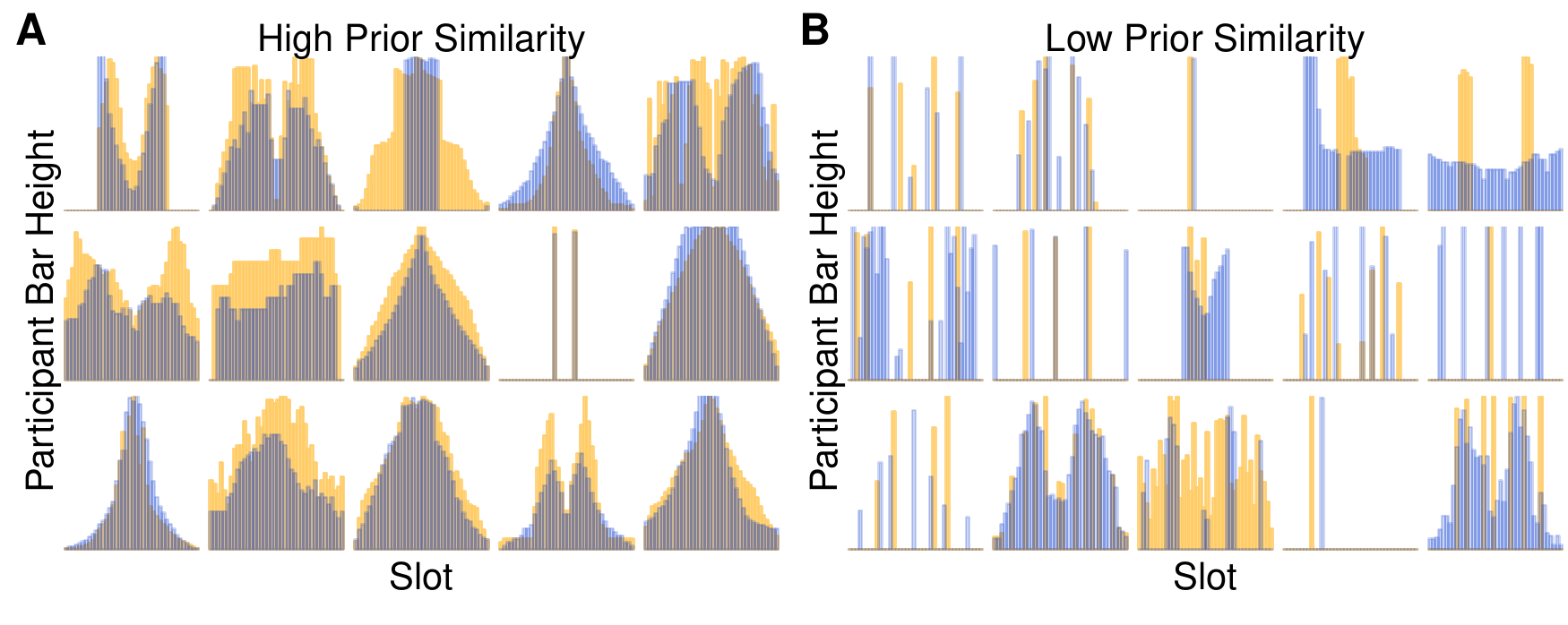} \caption[A]{A: Participant first (yellow) and second (blue) priors with a similarity measure greater than 0.50. Most prior pairs are similar in shape and are not jagged. B: Participant first (yellow) and second (blue) priors with a similarity measure less than 0.50. Most prior pairs contain at least one jagged prior, usually the first.}\label{fig:highLowSimAPPriors}
\end{figure}

\end{knitrout}

Our mathematical characterization of similarity is sensitive to slight shifts in sparsely drawn jagged probability distributions, resulting in low similarity ratings. Generally, when participants give a non-jagged prior, they tend to stick with it when asked to reproduce their priors. In contrast, sparsely drawn priors are unstable over time. The reason for this is unknown, but may involve the participants' confidence in their priors (given the result above), eagerness to begin the task, or a reconsideration of task instructions or goals.

These results are necessary but not sufficient to demonstrate reliability of prior elicitation. A more appropriate test of reliability should include a time interval between test and retest on the scale of hours or days, not minutes as we have in our data. We have attempted to maximize the utility of this test-retest by subjecting participants to an unexpected and unpleasant distraction task between the two instances of prior elicitation. Our finding here suggest good reason to invest in a more logistically challenging reliability analysis in future work.

The goal of prior reliability is more broad than the goal of clustering priors in Experiment 1. In this experiment, we directly test the overall similarity between each prior for each participant (accounting for both shape and central tendency) in order to explore the reliability of our method to elicit priors.  Alternatively,  measuring how many participants remain in their original cluster after independently clustering both the first and second recorded prior, would be appropriate for examining the reliability of the particular clustering method in question. We recommend this approach in future work, where priors are elicited twice (as in this experiment) over larger sample sizes (as in Experiment 1).

Repeated measurement of overall similarity between a participant's priors also allows for investigation of how priors are internally represented. For example, are priors actually selected from a higher-order distribution of distributions \citep{franke2016does, herbstritt2019complex}? Our preliminary findings of prior reliability suggest that the sampling method from such a `hyper-prior' may be rather narrow, if this hypothesis is true.

\subsubsection{Participant ball drop predictions approach uniform equilibrium}

Participant estimates are more similar to the theoretically expected uniform distribution at the final trial ($M =  0.74$, $SD = 0.19$) than at the initial trial ($M =  0.50$, $SD = 0.23$), $t(29) = 6.25), p  < .001$ (Figure \ref{fig:apFirstLastComboPlot}).

\begin{knitrout}
\definecolor{shadecolor}{rgb}{0.969, 0.969, 0.969}\color{fgcolor}\begin{figure}
\includegraphics[width=\textwidth,height=0.87\textheight,keepaspectratio]{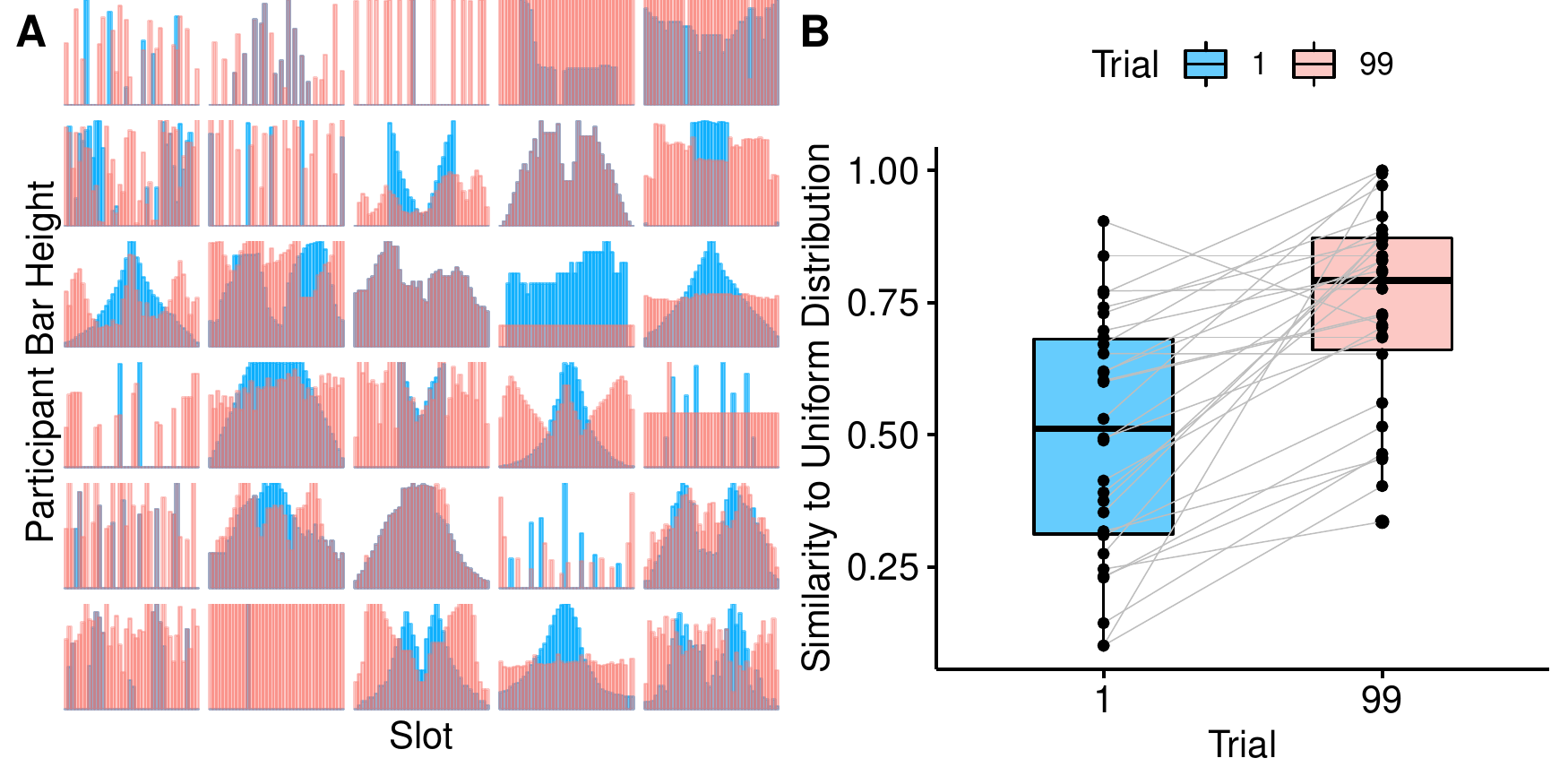} \caption[A]{A: Participant distribution estimates after the first (blue) and final (red) trials. Most participants become more similar to the uniform distrition, which is the theoretically expected equilibrium given our task construction. B: Participant estimate similarity to the theoretical equilibrium uniform distribution is greater on final trial (red) than first trial (blue).}\label{fig:apFirstLastComboPlot}
\end{figure}

\end{knitrout}

\begin{knitrout}
\definecolor{shadecolor}{rgb}{0.969, 0.969, 0.969}\color{fgcolor}\begin{figure}
\includegraphics[width=\textwidth,height=0.87\textheight,keepaspectratio]{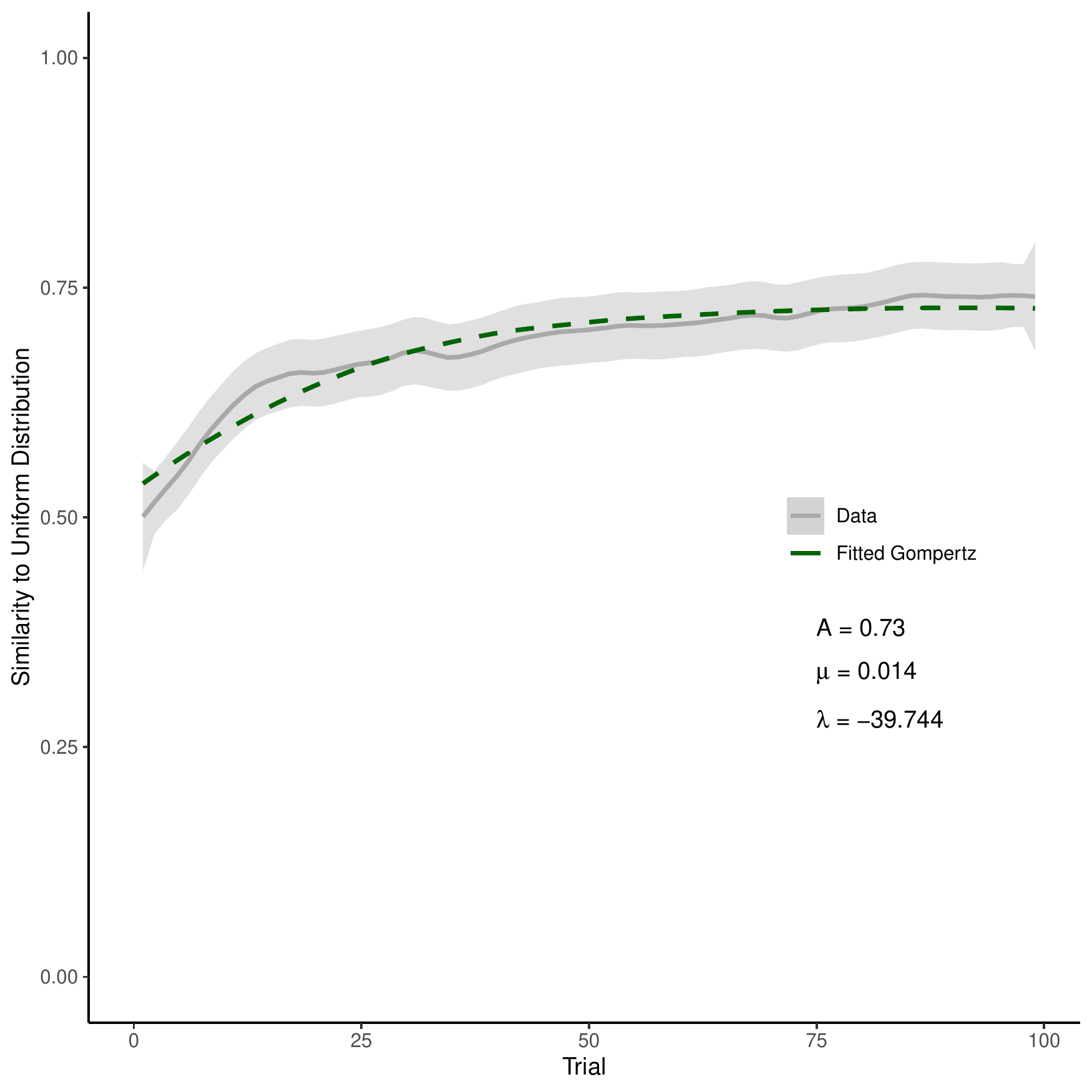} \caption[Aggregate participant estimation to uniform reference distribution +/- 95$\%$ CI with fitted Gompertz growth function]{Aggregate participant estimation to uniform reference distribution +/- 95$\%$ CI with fitted Gompertz growth function. The parameter $A$ defines the maximum similarity reached, $\mu$ defines maximum increase in similarity, and $\lambda$ defines the trial where $\mu$ occurs. Our fitted parameters indicate participant learning is gradual, integrating observed data continuously rather than intermittently.}\label{fig:avgAntiCurve}
\end{figure}

\end{knitrout}

Participants exhibit gradual learning, integrating observed trial data continuously rather than intermittently. Figure \ref{fig:avgAntiCurve} plots the aggregate data of all participants, with a fitted Gompertz growth function. Most informative are the fitted values of $\mu = 0.01$ and $\lambda = -39.74$. A small $\mu$ value indicates that aggregate participant behaviour contains no sharp increases in prediction accuracy. A negative $\lambda$ implies the inflection point of the fitted model occurs before the first trial, meaning the learning rate of our participants is monotonically decreasing as they approach the uniform equilibrium.

\subsubsection{The influence of true physics is negligible}

Participants' initial priors did not reflect the distribution expected if this were a physical game of Plinko. The similarity between participants' initial prior and the physically plausible physical distribution ($M = 0.33$, $SD = 0.24 $) is no different than the similarity between participants' first prior and 1000 randomly paired second priors ($M = 0.33$, $SD = 0.02$), $t(29) = -0.06$, $p  = .951$. This suggests that participant priors are no more similar to the physically plausible binomial distribution (had this been a physical game of Plinko) than randomly paired priors. Also, participants' \emph{explicit} estimations of the distribution of ball drops had this been a physical game of Plinko were no more accurate ($M = 0.33$, $SD = 0.18$) than the similarity between participants' first prior and 1000 randomly paired second priors ($M = 0.33$, $SD = 0.02$), $t(29) = -0.07$, $p  = .942$. Again, this suggests that participant estimates of a physical ball dropping through physical pegs are no more similar than randomly paired priors. 

Intuitive physics also does not inhibit participants' ability to learn physically implausible probability distributions. Participants' ball drop estimations on the final trial were more similar to the uniform distribution ($M = 0.74$, $SD = 0.19$) than to the distribution expected had the game been played with a physical ball and pegs ($M = 0.31$, $SD = 0.09$), $t(29) = 15.17$, $p  < .001$.

Our results suggest that Plinko is an effective tool for studying the priors and learning patterns of probabilistic representations in humans. Moreover, our results may not be limited to human perception of physical balls dropping through an array of pegs, since participant priors and explicit predictions of a physical game of Plinko do not resemble the physically plausible binomial distribution, and participants are able to effectively learn a distribution that is not physically plausible (the uniform distribution).

\section{General Discussion}

Here we present a task that provides a detailed representation of participant beliefs in a dynamic learning environment. We demonstrate the effectiveness of this task in three statistical learning experiments. In Experiment 1, we explored how participant priors indicate learning ability via three prior clustering methods. These results highlight the importance of measuring individual priors rather than assuming or retroactively inferring priors to apply uniformly to an entire group. Plinko provides a convenient avenue with which to measure individual participant priors. Matters of perceptual averaging and mental model smoothing can be examined by varying the distribution of present ball drops \emph{and} the chosen ``reference'' distribution with which to compare participant estimates. Future work should determine whether this smoothing is merely a function of drawing probability estimates, or truly represents a function of statistical learning and mental model updating, and to what limits this function can be pushed.

Experiment 2 demonstrated that participants are able to learn and represent a number of different probability distribution types. Participants were able to update their estimates when the ball drop distribution changed at unannounced points throughout the task, and this ability to update can be manipulated though experimental features such as breaks. Plinko is therefore an effective tool to examine both the influence of participant prior beliefs on statistical learning, and how effectively mental models can be updated when faced with new contingencies, given the nature of the task's environment. Future work using this task should explore the role participant priors play in updating to new task contingencies. 

In Experiment 3, we explored whether priors collected by our task are reliable and meaningful, since participants reproduce a similar prior after being told their original prior was lost and completing an ostensibly unrelated task. We also verified that intuitive physics of a literal Plinko game are not represented in participant behaviour, as priors did not resemble a physically plausible distribution, and participants were able to effectively learn a physically implausible distribution. This implies that our task can measure features of statistical learning that generalize beyond the stimulus-specific context of ball drops, since participants are not equally entrenched in the one `correct' physically plausible prior. We suspect this is either due to a lack of knowledge of literal Plinko physics, or a suspension of literal physical expectations in this computerized task, akin to how we are not surprised by superhero flight in a video game despite holding the prior belief that humans cannot fly.

For each experiment, we used angular similarity of participants' estimates represented as Euclidean Vectors to define learning accuracy to a reference distribution. Despite the benefits of scale invariant similarity measures, we do not commit to our selection as being the single best option. Our results from Experiment 3 demonstrate that our measure is particularly sensitive to lateral shifts, especially when comparing jagged probability distributions. This is true of any measure that assumes a literal interpretation of slot indices - something humans are unlikely to do. However, as seen in Experiment 3, remarkably jagged priors are 1) rare, and 2) may just represent a participant's lack of confidence in any particular shape of prior. In this case, the concern of a similarity measure's sensitivity to lateral shifts is mitigated since the most problematic use case is also the case where we are least interested in the literal shape of the participant's provided prior.

Conceptually, our similarity measure assumes that participant estimates represent what participants were instructed to represent - that the relative heights of the bars are proportionate to the relative probabilities of where the ball will land when dropped. This is likely the safest assumption to make in the absence of any research indicating how humans express internally represented probability distributions with computer mouse-drawn or touchscreen-drawn histograms. Regardless, future users of our Plinko task can elect to circumvent these issues by restricting the total area of participant ball drop estimates (thus removing the need for theoretical motivation of estimate normalizing). However, restricting participant estimates to sum to a normalized value may negatively impact usability, preventing the participant from providing maximally accurate representation of belief. Alternatively, if future work elects to both maximize degrees of freedom for participant responses, \emph{and} use a similarity measure that requires normalization, some consideration for the appropriate normalization method is required. For example, it may be that participant beliefs are actually best represented by the absolute, rather than relative differences between their drawn histogram bar heights. If this were true, subtractive normalization (subtracting an identical normalizing value from each bar height), rather than multiplicative normalization (multiplying each bar height by an identical normalization value) may be more appropriate \citep{rigoli2016neural}.

The terms ``mental models'' and ``statistical learning'' are generally used in reference to particular theoretical frameworks.
``Mental models'' traditionally refer to non-probabilistic rules or structures that inform human understanding and reasoning \citep{johnson1983mental, gentner1983mental}, though the utility of a probabilistic treatment has been demonstrated \citep{filipowicz2016, danckert2012, shaqiri2013statistical}. Here, we appeal to the more probabilistic interpretation of mental models. ``Statistical learning'' in Psychology traditionally refers to the ability to detect statistical regularities in one's environment. There is also considerable discussion as to the implicit and incidental nature of statistical learning \citep{perruchet2006implicit, arciuli2017multi,christiansen2019implicit}. Plinko is fundamentally a statistical learning task. Participants learn what event is likely to occur next, given some previously established understanding. This is as true for co-occurrences of syllables in a stream of speech as it is for co-occurrences in a series of ball drops. Our participants were explicitly asked to give predictions about the underlying model of the environment, rather than observed for indicators of implicit learning like reaction time. There are important considerations as to what particular kind of statistical learning tasks like these are actually measuring \citep{christiansen2019implicit}. As such, we leave these concerns for future work, and instead employ a broader interpretation of statistical learning that excludes concerns of the implicit and incidental nature of the traditional notion of statistical learning.

Ball drops are a convenient story to ascribe sequential discrete events for the purpose of measuring statistical learning, but our Plinko task is not limited to such a narrative. Visually presented ball drops are easier to administer and easier for participants to interpret than more abstract domains presented as numerical data points or other modalities. We believe ball drops are an ideal proof-of-concept stimulus to demonstrate the utility of our methodology. A generalized version of Plinko that does away with ball drops still affords researchers the ability to measure individual differences in prior belief structures, and how these beliefs are influenced by new events. Ball drops may be replaced by any other sequence of discrete events that could be easily mapped to a uni-dimensional ordered spatial domain over which a histogram can be drawn. For example, if interested in movie run-times \citep{griffiths2006optimal}, the bins of the Plinko histogram can represent a range of run-times, and individual ball drops can be replaced with presented instances of individual movie run-times. Individual priors of movie run-times, and the continuous updating thereof, can be analyzed in the same way we have demonstrated here. Other examples include height, age, color, grades, phonemes, and monetary values. 

If we believe people are learning probability distributions, we ought to measure them as directly and precisely as possible, while minimizing any preconceived notions of how probability ``should'' be represented. This means  maximizing degrees of freedom for participant responses and doing away with assumptions of Bayesian models, and families of well-defined parameterized probability distributions. Plinko can be used to refine our understanding about the individual differences of priors and how the contextual elements of a task affect our ability to revise prior beliefs. By adapting cognitive models to account for these factors, Plinko can contribute to a better understanding of human learning and updating.

\section{Acknowledgments}
We thank Caidence Paleske for assisting with data collection and analysis, and Andriy Struk and Jhotisha Mugon for including Plinko in their set of touchscreen foraging experiments.

\section{Funding}
JD and BA were each funded by NSERC Discovery awards. AF was funded by an NSERC Graduate Fellowship.

\section{Conflict of interest}

The authors declare that they have no conflict of interest.

\section{Open practices statement}

None of the experiments were preregistered. Data from Experiment 2 was previously reported in author AF's thesis \citep{filipowicz2017adapting}. Three \emph{other} Plinko experiments were run in the same time frame as the three presented here \citep{filipowicz2017adapting, filipowicz2018rejecting}. We did not perform any additional analysis that is not presented here. The data and materials for all experiments are available at https://osf.io/dwkie.

\bibliographystyle{apalike}

\bibliography{ms} 
\addcontentsline{toc}{section}{References}

\newpage

\section{Appendix: Measuring Similarity}

We measure similarity between two discrete probability distributions by computing the angle between each representative Euclidean vector of said distributions. Consider two probability distributions $P$ and $Q$ defined over a discrete random variable, $X$:\\

\begin{center}
$\begin{array}{c|cccc} x &1 &2 &\dots &n\\ \hline P(x) &p_1 &p_2 &\dots &p_n\\ \end{array}$  \hspace{1cm} $\begin{array}{c|cccc} x &1 &2 &\dots &n\\ \hline Q(x) &q_1 &q_2 &\dots &q_n\\ \end{array}$\\
\end{center}

Distributions $P$ and $Q$ are uniquely represented by the vectors $\vec{p}$ and $\vec{q}$, respectively:

  \begin{align*}
    \vec{p} &= \begin{bmatrix}
           p_{1} \\
           p_{2} \\
           \vdots \\
           p_{n}
    \end{bmatrix} \hspace{1cm}
        \vec{q} = \begin{bmatrix}
           q_{1} \\
           q_{2} \\
           \vdots \\
           q_{n}
         \end{bmatrix}
  \end{align*}

Any discrete probability distribution, and any histogram a participant may draw in our experiments, is restricted to only contain non-negative values. Therefore, any arbitrary $\vec{p}$ and $\vec{q}$ exist in the positive orthant of Euclidean space, denoted ${\rm I\!R}^{n}_+$. Euclidean space affords a notion of an angle $\theta$ (in radians) between any non-zero vectors $\vec{u}$ and $\vec{v}$, defined as:

\begin{equation}
\theta(\vec{u}, \vec{v}) = \arccos  \left( \frac{\vec{u} \cdot \vec{v}}{ \left\Vert \vec{u} \right\Vert \left\Vert \vec{v} \right\Vert} \right)
\end{equation} where $\vec{u} \cdot \vec{v} = \sum\limits_{i=1}^{n}{u_i}{v_i}$, and $\left\Vert \vec{u} \right\Vert = \sqrt{\vec{u} \cdot \vec{u}}$. Since $\vec{p}$ and $\vec{q}$ are restricted to ${\rm I\!R}^{n}_+$, $\theta(\vec{p}, \vec{q}) \in [0,\frac{\pi}{2}]$. When a distribution $P$ is identical to (or a scalar multiple of) $Q$, the angle between $\vec{p}$ and $\vec{q}$ is 0. When a distribution $P$ shares no mutual slot mass with distribution $Q$, the angle between $\vec{p}$ and $\vec{q}$ is $\frac{\pi}{2}$ ($90 ^{\circ}$). For easier interpretation, a linear transformation is applied to $\theta(\vec{p}$, $\vec{q})$ to create a measure of similarity, denoted $S(\vec{p}, \vec{q})$, where

\begin{equation}
S(\vec{p}, \vec{q}) = 1 - \frac{2}{\pi} \theta(\vec{p}, \vec{q})
\end{equation}

This measure of similarity ranges from 0 when $\vec{p}$ and $\vec{q}$ are maximally dissimilar, to 1 when $\vec{p}$ and $\vec{q}$ are maximally similar. A fully algebraic definition of $S$ combines (1) and (2):

\begin{equation}
S(\vec{p}, \vec{q}) = 1 - \frac{2}{\pi} \arccos \left( \frac{\sum\limits_{i=1}^{n}{p}_{i}{q}_{i}}{\sqrt{\sum\limits_{i=1}^{n}{p}_{i}^{2}} \sqrt{\sum\limits_{i=1}^{n}{q}_{i}^{2}}} \right) 
\end{equation}

\subsection{A proof of scale invariance}

In our paper, we claim that $S(\vec{p}, \vec{q})$ is invariant to the scale or multiplicative normalization of any given $\vec{p}$ or $\vec{q}$. Intuitively, changing the scale (length) of two vectors does not change the angle between them. We demonstrate this algebraically:

\vspace{0.5cm}

Consider  $\vec{p}, \vec{q} \in {\rm I\!R}^{n}_+$ and $a, b \in {\rm I\!R}$ such that $a, b \neq 0$. Then,

\begin{align}
  S(a\vec{p}, b\vec{q}) &= 1 - \frac{2}{\pi} \arccos \left( \frac{\sum\limits_{i=1}^{n}{ap}_{i}{bq}_{i}}{\sqrt{\sum\limits_{i=1}^{n}a^2{p}_{i}^{2}} \sqrt{\sum\limits_{i=1}^{n}b^2{q}_{i}^{2}}} \right)\\
  &= 1 - \frac{2}{\pi} \arccos \left( \frac{ab\sum\limits_{i=1}^{n}{p}_{i}{q}_{i}}{\sqrt{a^2\sum\limits_{i=1}^{n}{p}_{i}^{2}} \sqrt{b^2\sum\limits_{i=1}^{n}{q}_{i}^{2}}} \right)\\
  &= 1 - \frac{2}{\pi} \arccos \left( \frac{ab\sum\limits_{i=1}^{n}{p}_{i}{q}_{i}}{\sqrt{a^2}\sqrt{\sum\limits_{i=1}^{n}{p}_{i}^{2}} \sqrt{b^2}\sqrt{\sum\limits_{i=1}^{n}{q}_{i}^{2}}} \right)\\
  &= 1 - \frac{2}{\pi} \arccos \left( \frac{ab\sum\limits_{i=1}^{n}{p}_{i}{q}_{i}}{ab\sqrt{\sum\limits_{i=1}^{n}{p}_{i}^{2}} \sqrt{\sum\limits_{i=1}^{n}{q}_{i}^{2}}} \right)\\
  &= 1 - \frac{2}{\pi} \arccos \left( \frac{\sum\limits_{i=1}^{n}{p}_{i}{q}_{i}}{\sqrt{\sum\limits_{i=1}^{n}{p}_{i}^{2}} \sqrt{\sum\limits_{i=1}^{n}{q}_{i}^{2}}} \right)\\
  &= S(\vec{p}, \vec{q})\\[1em]
  \therefore \hspace{0.5cm} S(a\vec{p}, b\vec{q}) &= S(\vec{p}, \vec{q}) \hspace{1cm} \square
\end{align}

\end{document}